\documentclass[
reprint,prd,aps,nofootinbib]{revtex4-1}

\usepackage{graphicx}
\usepackage{dcolumn}
\usepackage{bm}
\usepackage{hyperref}
\usepackage{amsmath}
\usepackage{amssymb}
\usepackage{amsfonts}
\usepackage{resizegather}
\usepackage{caption}

\usepackage{color}
\usepackage{ulem}
\usepackage{cancel}
\usepackage[export]{adjustbox}
\usepackage[toc,page]{appendix}

\newcommand{\lmatt}{\mathcal{L}_\text{matt}}

\begin{document}

\title
{Non-vacuum relativistic extensions of MOND using metric theories of gravity with curvature-matter couplings and their applications to the accelerated expansion of the Universe without dark components}

\author{Ernesto Barrientos}
\email{ernestobar14@ciencias.unam.mx}
\affiliation{Instituto de Investigaciones en Matem\'aticas Aplicadas 
	         y en Sistemas, Departamento de Matem\'aticas y
		 Mec\'anica, Universidad Nacional Aut\'onoma de M\'exico,
		 Ciudad de M\'exico 04510, M\'exico.}
\author{Tula Bernal}
\email{tbernalm@chapingo.mx}
\affiliation{Universidad Aut\'onoma Chapingo, km. 38.5 Carretera M\'exico-Texcoco, Texcoco 56230, Estado de M\'exico, M\'exico}
\author{Sergio Mendoza}
\email{sergio@astro.unam.mx}
\affiliation{Instituto de Astronom\'{\i}a, Universidad Nacional Aut\'onoma de M\'exico, AP 70-264, Ciudad de M\'exico 04510, M\'exico}

\date{\today}


\begin{abstract}
  We discuss the advantages of using metric theories of gravity
with curvature-matter couplings in order to construct a relativistic
generalisation of the simplest version of Modified Newtonian Dynamics
(MOND), where Tully-Fisher scalings are valid for a wide variety of
astrophysical objects.  We show that these proposals are valid at the
weakest perturbation order for trajectories of massive and massless
particles (photons).  These constructions can be divided into local and
non-local metric theories of gravity with curvature-matter couplings.
Using the simplest two local constructions in a FLRW universe for dust,
we show that there is no need for the introduction of dark matter and
dark energy components into the Friedmann equation in order to account
for type Ia supernovae observations of an accelerated universe at the present 
epoch.
\end{abstract}

\pacs{04.50.Kd, 04.20.Fy, 95.30.Sf, 98.80.Jk,04.25.-g, 04.20.-q}

\maketitle

\section{Introduction}
\label{introduction}

The concordance cosmological model requires the addition of two dark
components into the energy-momentum tensor in order to balance the
Hilbert-Einstein field equations, i.e.~in order that Einstein's
tensor is \( 8 \pi G / c^4 \) times the energy-momentum tensor,
with \( c \) the speed of light and \( G \) Newton's constant of
gravity.  These  dark matter and dark energy components  are in very
good agreement with a wide variety of cosmological observations
\citep[see][among others]{White93,Ostriker95,Riess98,Perlmutter99,Tegmark2004,Bertone:2004,Cole2005,Eisenstein2005,Komatsu2011,Bennett:2013,planck}.
However, the lack of any direct or indirect detection of dark matter and
the more complicated understanding of the nature of the dark energy, have
opened up different paths to explore this issue in the last decades.

One such exploration path, which is the one 
we are going to use in this article, is to assume
that the balance of the Hilbert-Einstein field equations requires
an extension or a modification. This line of research
might poses more profound implications into how weak or strong the presence of matter and energy in 
space is related with the curvature of space-time, a fact that by itself  must be
investigated.  The immediate generalisation consists in 
building a more general Einstein tensor through
a generalisation of the  Hilbert action \citep[see e.g.][]{landau-fields}:
\begin{equation}
  S = - \frac{ c^3 }{ 16 \pi G } \int{ R \, \sqrt{ -\textsl{g} } \, \mathrm{d}^4 x } - \frac{ 1 }{ c }
  \int{ \mathcal{L}_\text{matt} \, \sqrt{-\textsl{g}} \, \mathrm{d}^4 x },
\label{eq001}
\end{equation}

\noindent by allowing for Ricci's scalar \( R \) to become a general
$f(R)$ function \citep{capozziello-book}:

\begin{equation}
  S = - \mathcal{K}  \int{ f(R) \, \sqrt{ -\textsl{g} } \, \mathrm{d}^4 x } - \frac{ 1 }{ c }
  \int{ \mathcal{L}_\text{matt} \, \sqrt{-\textsl{g}} \, \mathrm{d}^4 x } ,
\label{eq002}
\end{equation}

\noindent where $\textsl{g}:=\det{[g_{\mu\nu}]}$ is the determinant of the metric tensor $g_{\mu\nu}$, and $\mathcal{K}$ is a coupling constant. Such kind of models
are the so-called $f(R)$\textit{-theories}.

  In the previous equations, \( \mathcal{L}_\text{matt} \) is the
standard matter Lagrangian, which is connected to the energy-momentum
tensor \(T_{\mu\nu}\) through the relation~\citep[see e.g][]{ landau-fields,harko-lobo-book,franklin17,nuastase19}:

\begin{equation} T_{\mu\nu} := - \frac{ 2  }{ \sqrt{-\textsl{g}} }
      \frac{ \delta \left( \sqrt{-\textsl{g}} \, \mathcal{L}_\text{matt} \right) }{ \delta
    g^{\mu\nu} } .
\label{energy-momentum-tensor}
\end{equation}

\noindent In what follows, we use a metric signature \((+,-,-,- )\).
Latin indices take space values \(1,2,3\), while Greek ones take
space-time values \(0,1,2,3\).  The first term on the right-hand side of
equation~\eqref{eq001} is Hilbert's gravitational action \( S_\text{H}
\), while the second term represents the standard matter action \(
S_\text{matt} \).

  The introduction of a general \( f(R) \) function 
into the gravitational action \( S_\textrm{g} \) (first term on 
the right-hand side of equation~\eqref{eq002}) produces fourth-order
differential field equations when null variations with respect to the
metric tensor are carried out 
\citep[see e.g.][]{capozziello-book}.

  In the non-relativistic regime, the first modification to Newton's gravity proposed in order to explain the rotation curves and the Tully-Fisher relation of galaxies was the Modified Newtonian Dynamics (MOND) \citep{Milgrom1,Milgrom2}.
The main idea of MOND is essentially an assumption that
Kepler's third law of planetary motion is not correct at all scales.  This is 
based on a wide range of observations from dynamical studies of binary stars,
spiral galaxies and dynamical pressure supported systems such as globular clusters, elliptical
galaxies and clusters of galaxies (\citet{Xavier_proc, Xavier_scales}). The key ingredient to understand is that for 
the case of circular orbits, Kepler's third law of planetary motion \( v \propto \sqrt{M} / r \) 
(where $v$ is the velocity or the velocity dispersion at radius \( r \) of a system with internal mass \( M \)) changes to a Tully-Fisher behaviour \( v \propto M^{1/4} \) with 
no dependence on the 
radial distance \( r \) at sufficiently large distances, where the equivalent Newtonian acceleration
reaches a value of  \( a_0 \approx 1.2 \times 10^{-10} {\rm m / s^2 } \).  For the case of a test particle
orbiting in circular motion about a point mass \( M \) (like a planet orbiting about the Solar System), Newton's basic 
formula for gravity is obtained using the fact that  the acceleration \( a \) exerted by gravity on this test
particle is given by the centrifugal force per unit mass: \( a= v^2 / r  \propto M / r^2 \).  Following the same 
procedure when Kepler's third law is changed by the Tully-Fisher law, then \( a = v^2 / r \propto
\sqrt{M} / r \), in the so-called ``deep-MOND'' regime.  
The constant of proportionality is calibrated with e.g. the dynamics of rotation curves in
spiral galaxies and can be written as \citep{mendoza15}\footnote{As in Newtonian gravity,
the constant of proportionality \( G_\text{M} \) in equation~\eqref{mond-basic} requires to 
be a fundamental constant of nature.  Its value is \( G_\text{M}  \approx
8.95 \times 10^{-11} {\rm m }^2 { \rm s }^{-2} {\rm kg }^{-1/2} \), but since MOND's basic
formula was originally constructed to be a modification to inertia, then it is tradition to 
introduce an ``ill defined'' MONDian acceleration \( a_0 := G_\text{M}^2 / G \) into the fundamental constants scenario and forget about \( G_\text{M} \) \citep{mendoza15}.}: 

\begin{equation}
  a = - G_\text{M} \frac{\sqrt{M}}{r} = - \frac{ \sqrt{  a_0 G M } } { r }. 
\label{mond-basic}
\end{equation}

  In this sense, the MONDian acceleration \( a_0 \)  serves as a way to choose the regime of 
gravity: the Newtonian regime or the deep MONDian regime or some transition function in between.
Modifications of gravity at the non-relativistic regime using this approach were constructed by
\citet{Mendoza:2010} and it was shown using theoretical grounds and calibrations to Solar System
dynamics that the transition zone occurred quite abruptly.  Using this approach in globular
clusters, \citet{hernandez12b} and \citet{hernandez} showed very precisely that 
the transition occurs in a very abrupt manner and so there can only be a full Newtonian regime 
or a deep MONDian regime with an abrupt transition between one and the other ocurring at accelerations 
\( a_0 \).

Moreover, in the search of a fundamental theory of MOND, Milgrom first noticed the coincidence relation: $2 \pi a_0 \approx c H_0$, with $H_0$ the Hubble constant at the present epoch \citep{Milgrom1}. The Hubble radius can be written as $R_H = c/H_0$ and the Hubble mass is given by $M_H = c^3/GH_0$. From these two relations we have $a \approx GM_H/R_H^2 \approx 2\pi a_0$ \citep[see][]{Milgrom:2008,Bernal:2011a}. Thus, the Newtonian gravitational acceleration at the present time is approximately MONDian. In this sense, we should develop a MONDian relativistic treatment to understand the dynamics of the Universe without dark components. Also, it has been proposed that MOND's acceleration constant is a fundamental constant that arises from astrophysical phenomenology \citep{Mendoza:2010,Bernal:2011a}.

A full potential non-relativistic theory of MOND was constructed by \citet{Bekenstein:1984}
using an AQUAdratic Lagrangian (AQUAL) which happens to predict an external field effect \citep{binney-famaey}.  Since an external field effect is not present in the galaxy with wide open
binaries motion not following Newton's law of gravitation \citep{hernandez12a,Hernandez:2019}, then one can 
no rely on all the results such approach present.

  \citet{Bekenstein} constructed a complicated relativistic theory using tensor, vector and 
scalar fields, which at the weak field limit of approximation converges to MOND.  The 
mathematical complexities of the theory and the many problems it had presented to fit
various astrophysical scenarios \citep[see e.g.][and references therein]{Ferreras:2009} required thoughts on
generalisations even at the most fundamental aspects of gravitation: the action.

In previous studies, it has been shown that in order to
build a relativistic version of MOND using
$f(R)$ theories in the pure metric approach or in the Palatini
formalism or even with the inclusion of torsion, curvature-matter
couplings need to be introduced into the gravitational action
\citep{Mendoza:2010,bernal11,mendoza13,barrientos16,barrientos17,barrientos18}. 
\citet{Bertolami} showed that for a particular generalisation of the $f(R)$ theories in the metric approach, by coupling the $f(R)$ function with the Lagrangian density of matter $\mathcal{L}_\text{matt}$, an extra-force arises which in the weak field limit can be connected with MOND's acceleration and possibly explain the Pioneer anomaly.

  Working with a static spherically symmetric space-time,
\citet{bernal11} constructed a gravitational \( f(R) \) action in
the pure metric approach which depends explicitly on the  central
point mass \( M \) of the problem. This model became very successful
explaining the non-relativistic Tully-Fisher weak field limit and turned
out to be coherent when explaining deflection of light on individual,
groups and clusters of galaxies \citep{mendoza13}.  The same model was
also applied to reproduce the dynamical masses of 12 \textit{Chandra}
X-ray galaxy clusters, through the fourth-order metric coefficients,
roughly extending the model to mass distributions \citep{Bernal:2019}.
Such theory was also able to describe a correct acceleration of the
local Universe, without the introduction of dark matter and dark energy
components, through type Ia supernovae (SNe~Ia) observations, for the
functional mass \( M \) taken as the causally connected mass to a given
fundamental observer moving with the Hubble flow, i.e.~it was taken to
be the Hubble mass \citep{carranza13}.

  Since the mass function \( M \) depends on the amount of matter
at any given point for a given fundamental observer, the action is
\textit{non-local} \citep{qft,nonlocal, Maggiore2014,Hehl2009,Blome}.
Although, in principle, there is nothing wrong with the construction of a
non-local theory of gravity, we have got adapted to the idea of locality
in physics, and gravitational theory is no exception.  The search of a
\textit{local} relativistic theory of gravity in the full MONDian regime with a
pure metric approach was carried on with success by \citet{barrientos18},
where the bending of light was successfully reproduced on such
regime.  Attempts on the construction of relativistic theories of
MOND in the metric-affine formalism and with torsion were carried
out by \citet{barrientos16} and \citet{barrientos17} respectively.
The metric-affine one being non-local.  In these two last cases, the
resulting equations are quite cumbersome and lead to no-simple paths
for their applicability in astrophysical and cosmological environments.

  In summary, all previous results signal that curvature-matter
couplings are required with a more general action:

\begin{equation}
  S = \int{ F(R, \mathcal{L}_\text{matt}) \, \sqrt{-\textsl{g}} \, \mathrm{d}^4 } x,
\label{matter-curvature}
\end{equation}
  
\noindent as described
in \citet{frlm,Bertolami,Allemandi}, all
generalisations of the pioneer work of \citet{Goenner:1984}.  In the
previous equation, \( F \) is a general function of the Ricci scalar \(
R \) and the matter Lagrangian \( \mathcal{L}_\text{matt} \).  The null
variations of action~\eqref{matter-curvature} with respect to the metric
tensor yield the following field equations \citep{frlm}: 

\begin{equation}
  \begin{split}
  F_R R_{\alpha\beta} + &  \left( g_{\alpha\beta} \nabla^\mu \nabla_\mu -
  \nabla_\alpha \nabla_\beta \right) F_R - \frac{ 1 }{ 2 } \left( F -
  \lmatt F_{\lmatt} \right)g_{\alpha\beta} \\ 
  & = \frac{ 1 }{ 2 } F_{\lmatt}
  T_{\alpha\beta},
  \end{split}
\label{fieldeqs}
\end{equation}

\noindent with $R_{\alpha\beta}$ Ricci's tensor and the trace given by:

\begin{equation}
  F_R R + 3 \Delta F_R - 2 \left( F - \lmatt F_{\lmatt} \right) = \frac{ 1 }{ 2 }
  F_{\lmatt} T,
\label{trace}
\end{equation}

\noindent where $T:=T^\alpha_\alpha$ is the trace of the energy-momentum tensor, $F_R:= \partial F/ \partial R$ and
$F_{\mathcal{L}_\text{matt}}:=\partial F / \partial
\mathcal{L}_\text{matt}$.  The motion of test free
particles is expressed through the following non-geodesic equation:

\begin{equation}
  u^\alpha \nabla_\alpha u^\beta = \frac{ \mathrm{D} u^\beta }{\mathrm{d}
    s }
    = \frac{ \mathrm{d}^2  x^\beta }{\mathrm{d} s^2 }  +
    \Gamma^\beta_{\ \mu\nu} 
    \frac{ \mathrm{d}  x^\mu }{\mathrm{d} s } 
    \frac{ \mathrm{d}  x^\nu }{\mathrm{d} s } = f^\beta,
\label{geodesic}
\end{equation}

\noindent with \( \Gamma^\alpha_{\mu\nu} \) representing the Levi-Civita
connection, \( \text{d} s \) the interval  and the curvature-matter extra 
force vector \( f^\beta \) is given by the following relation:

\begin{equation}
  f^\beta :=  \left( g^{\beta \lambda } - u^\beta u^\lambda \right)
  \nabla_\lambda \left( F_\mathcal{L_\text{matt}}  
  \frac{ \mathrm{d}  \mathcal{L}_\text{matt} }{\mathrm{d} \rho }\right),
\label{force}
\end{equation}

\noindent where $\lmatt$ is the matter Lagrangian which will be further discussed 
in Section \ref{generalities}. The extra force \( f^\beta \) generated by the curvature-matter
coupling is orthogonal to the velocity \( u^\beta \), since \( u_\alpha
\nabla_\beta u^\alpha = 0 \).

  In this article, we show that by using the pure metric approach of an
extended gravitational action with curvature-matter couplings it is possible to
build an infinite number of non-local and local
relativistic descriptions of MOND as their weak-field limit.
Since the interpretation of the mass function \( M \) might be only evident
in systems with high degrees of symmetry (such as  spherically symmetric
ones and/or isotropic spaces), we chose what appears to
be the two simplest proposals and test their validity in an expanding
Friedmann-Lema\^itre-Robertson-Walker (FLRW) Universe, with no dark
matter, nor dark energy components.

   In Section~\ref{generalities} we describe the simplest assumptions we
can make for the Lagrangian and show that there are an infinite number of
relativistic theories that can yield MOND in the weak-field regime. In
Section~\ref{cosmography} we discuss the use of the cosmographic
parameters as a model-independent way to constrain the cosmological
observables. In Section~\ref{cosmo} we show the field equations in
a cosmological scenario using the FLRW metric with curvature-matter
couplings, in view of the constrains to the cosmographic parameters
for two specific models. In Section~\ref{ajuste} we show the results
of fitting the two models chosen with SNe~Ia observations.  Finally,
in Section~\ref{discussion} we discuss our results and present our
conclusions.

\section{General metric curvature-matter coupling}
\label{generalities}

The field equations \eqref{fieldeqs} for a general
function $F(\chi,\xi)$, where $\chi := \alpha R$ and $\xi := \lmatt /
\lambda$, with $\alpha$ and $\lambda$ coupling constants which make $\xi$
and $\chi$ dimensionless quantities, are given by \citep{barrientos18}:

\begin{equation}
	\alpha F_\chi R_{\mu\nu} -\frac{1}{2}g_{\mu\nu}(F-F_\xi \xi)+\alpha(g_{\mu\nu}\Delta-
	\nabla_\mu\nabla_\nu)F_\chi
	=\frac{F_\xi T_{\mu\nu}}{2\lambda} ,
	\label{field.eqs.}
\end{equation}

\noindent with $F_\chi := \partial F / \partial \chi$, $F_\xi := \partial F / \partial \xi$ and the Laplace-Beltrami operator $\Delta:=\nabla^\mu \nabla_\mu$.
The last equation can be 
written as

\begin{equation}
	R_{\mu\nu} -\frac{1}{2}g_{\mu\nu}R=\frac{1}{\alpha F_\chi}\left[\frac{F_\xi T_{\mu\nu}}{2\lambda}
	-\frac{1}{2}g_{\mu\nu}F_\xi \xi +T_{\mu\nu}^\text{curv}\right],
	\label{field.eqs.eins}
\end{equation}

\noindent where we have defined

\begin{equation}
	T_{\mu\nu}^\text{curv} := \frac{1}{2}g_{\mu\nu}(F-\chi F_\chi)-\alpha \left( g_{\mu\nu}\Delta
	-\nabla_\mu \nabla_ \nu \right) F_\chi.
	\label{Tcurv}
\end{equation}

  In order to construct a relativistic theory which converges to 
MOND in its weak-field limit, we follow the procedures shown in 
\citet{bernal11,barrientos17,barrientos18}, and as such we 
assume that the function $F(\chi,\xi)$ has the following form:

\begin{equation}
  F(\chi,\xi) = \chi^p \xi^u + \xi^v,
\label{frgeneral}
\end{equation}

\noindent where $p$, $u$ and $v$ are unknown real numbers. 

To build the coupling constants $\alpha$ and $\lambda$, we use
Buckingham's $\Pi$ theorem of dimensional analysis \citep[cf.][]{sedov}
in the following way. Since we are interested in a relativistic action
for MOND, we choose $c$, $G$ and $a_0$ as our independent variables. The
key feature of MOND is the introduction of a fundamental acceleration
constant $a_0 = 1.2 \times 10^{-10} \text{ms}^{-2}$, as a cut-off
scale into gravitational phenomenae \citep[see e.g.][]{Mendoza:2010,mendoza15}.
The dimensions of these three constants are given by

\begin{equation}
    [c]=lt^{-1}, \quad [a_0]=lt^{-2} \quad \mbox{and} \quad [G]=l^3t^{-2}m^{-1},
    \label{dimensiones}
\end{equation}

\noindent where $l$, $t$ and $m$ stand for dimensions of  length, time
and mass, respectively. It is straightforward to show that it is not
possible to write $c$, $G$ and $a_0$ as a combination of the other two and
therefore they are useful choices to take as independent dimensional
variables.  Since the dimensions of $R$ and $\lmatt$ are

\begin{equation}
    [R]=l^{-2}\qquad \mbox{and} \qquad [\lmatt]=l^{-1}mt^{-2},
    \label{dim2}
\end{equation}

\noindent then by demanding that

\begin{equation}
    [R]=[c]^a[G]^b [a_0]^d \quad \mbox{and} \qquad [\lmatt]=[c]^A[G]^B[a_0]^D,
\end{equation}

\noindent we obtain $a=-4$, $b=A=0$, $d=2$, $B=-1$ and $D=2$. Thus,
we can see that the coupling constants $\alpha$ and $\lambda$ are given by

\begin{equation}
    \alpha=k\frac{c^4}{a_0^2} \quad \mbox{and} \quad \lambda=k'\frac{a_0^2}{G} ,
    \label{constantes}
\end{equation}

\noindent where $k$ and $k'$ are constants of proportionality.

  The matter is described by a perfect fluid for which its energy-momentum tensor
is given by \( T_{\mu\nu} = \left( e +p \right)u_\mu u_\nu -p g_{\mu\nu} \), where $u_\mu$ 
is the normalised four-velocity with: $u^\mu u_\mu = 1$, with total energy density (rest mass density plus internal energy density) \( e \) and pressure \( p \) \citep{landau-fields}. The 
total energy density takes the value \( e = \rho c^2 \) for a dust dominated universe with \( p = 0 \), and \( e = ( \kappa - 1 ) p \) for a
radiation dominated Universe whit \( \kappa = 4/3 \) and a cosmological vacuum dominated Universe whit \( \kappa = 0 \)
\citep[see e.g.][]{longair-galaxy-formation}.  In the present article we are only interested in cosmological applications for a dust dominated 
universe and so:

\begin{equation}
T_{\mu\nu} = \rho c^2 u_\mu u_\nu.
\label{Tpolvo}
\end{equation}


  The matter Lagrangian for dust is given by~\citep{mendoza21}:

\begin{equation}
  \lmatt =  \rho c^2
\label{lmatt} 
\end{equation}

  Throughout this article, we will work with a dust dominated Universe.
Therefore, we will use the energy-momentum tensor of a
perfect fluid with pressure $p=0$  which leads to relation~\eqref{Tpolvo} and equation~\eqref{lmatt}.

  To order of magnitude, the trace of the field equations \eqref{field.eqs.}
takes the following form:

\begin{equation}
     F_\chi \chi +(F-F_\xi \xi)+\alpha\Delta F_\chi \sim \frac{F_\xi T}{\lambda}.
	\label{trace2}
\end{equation}

\noindent Using function \eqref{frgeneral}, the last expression turns into:

\begin{equation}
    \chi^p \xi^u + \xi^v + \alpha\Delta (\chi^{p-1} \xi^u) \sim  \frac{(\chi^p\xi^{u-1}+\xi^{v-1})T}{\lambda}.
	\label{trace3}
\end{equation}

\noindent For the case of dust $T=\lmatt$ and so, the last equation is 
expressed as:

\begin{equation}
    \alpha\Delta (\chi^{p-1} \xi^u)  \sim \chi^p\xi^u+\xi^v.
	\label{trace4}
\end{equation}

  We now explore the previous equation to perturbation order of powers
of $c^{-1}$ (order $\mathcal{O}(1)$)\footnote{We use the notation
$\mathcal{O}(n)$ for $n=0,1,2,...$ meaning $\mathcal{O}(c^0)$,
$\mathcal{O}(c^{-1})$, $\mathcal{O}(c^{-2}),...$, respectively,
following the well known notation of \citet{Will}.}. To do so, note
that the metric coefficients at second perturbation order are given
by \citep[see e.g.][]{sergioyolmo}\footnote{In general terms, when using
isotropic coordinates,  the space
components of the metric \( g_{ij} = \delta_{ij}\left(-1 + 2 \gamma 
\phi/ c^2 \right) \), but as shown by \citet{mendoza13,mendoza15,sergioyolmo}
the first Parametrised Post Newtonian (PPN) Parameter \( \gamma = 1 \)
when studing the bending of light of individual, groups and clusters
of galaxies under the assumption of a non-relativistic scalar MONDian 
potential.}:

\begin{equation}
  \begin{split} 
  g_{00} &= {}^{(0)}g_{00} + {}^{(2)}g_{00}=1+\frac{2\phi}{c^2},\\
  g_{ij} &= {}^{(0)}g_{ij} + {}^{(2)}g_{ij}=
  \delta_{ij}\left(-1+\frac{2\phi}{c^2}\right), \\
  g_{0i} &= 0,
  \end{split}
\label{metricaperturbada}
\end{equation}

\noindent where $\phi$ is the non-relativistic scalar gravitational
potential. There is not need to expand the metric beyond a second order in
powers of $c^{-1}$ since in the weak field limit of a relativistic theory,
the dynamics of massive particles is determined by the $\mathcal{O}(2)$
time-component of the metric, while the deflection of light is determined
by the $\mathcal{O}(2)$ radial one \citep{Will,Will:2006}.

  With the metric \eqref{metricaperturbada}, the Ricci scalar is given by:
$R=-2\nabla^2\phi/c^2$. Using these results and the coupling constants
given in \eqref{constantes}, the perturbation orders
of the terms in  equation \eqref{trace4} are given by:

\begin{equation}
    \overbrace{\alpha\Delta (\chi^{p-1} \xi^u)}^{{\cal{O}}(2(p+u+1))} \sim 
    \overbrace{\chi^p\xi^u}^{{\cal{O}}(2(p+u))}+\overbrace{\xi^v}^{{\cal{O}}(2v)}.
	\label{trace5}
\end{equation}

\noindent Since $(p+u+1)>(p+u)$, the coefficients are choose
to satisfy the relation $p+u+1=v$. Using this and, since to order of
magnitude, $\nabla\sim r^{-1}$, the trace \eqref{trace4} of the field
equations can be expressed as:

\begin{equation}
    \frac{\alpha^p R^{p-1}\lmatt^u}{r^2 \lambda^u}\sim \frac{\lmatt^v}{\lambda^v}.
    \label{aprox}
\end{equation}

\noindent To order of magnitude, the acceleration $a=|\nabla\phi|\sim
\phi/r$, and for a point mass $M$ the matter Lagrangian
\eqref{lmatt} takes the form \( \lmatt = \rho c^2 \sim
M/r^3 \). With these results we obtain:

\begin{equation}
    \frac{a^{p-1}c^{2(p+1)}}{r^{p+1}a_0^{2p}} \sim \frac{(MG)^{v-u}c^{2(v-u)}}{a_0^{2(v-u)}r^{3(v-u)}} \sim \left( \frac{MGc^2}{a_0^2 r^3}\right)^{v-u}.
\label{localac}
\end{equation}

\noindent The powers of $c$ must be the same in both sides
of last equation. This is achieved when $v=p+u+1$ and so, 
the expression for the acceleration takes the following form:

\begin{equation}
    a=\frac{(MG)^{(p+1)/(p-1)}}{a_0^{2/(p-1)}r^{2(p+1)/(p-1)}}.
    \label{aprox2}
\end{equation}

  The acceleration in the so-called ``deep-MOND'' regime is
$a=\sqrt{MGa_0}/r$ \citep{Milgrom1, Milgrom2}.  In order to obtain the
correct power for all the variables involved in the last equation ($M$, $G$,
$a_0$ and $r$) and the MOND-like limit, the value $p=-3$ is found. For
this value, the relation between $v$ and $u$ is $v=u-2$. Since there
is not another constriction, we conclude that there exist an infinite
number of models which yield the MONDian basic formula.

Amongst all those
models, we are interested in two cases: the first one was introduced by
\citet{barrientos18}, where the authors were looking for an action with
the matter Lagrangian alone in the matter sector, as is traditionally
done, i.e.~$v=1$, in consequence $u=3$ and so $ F(\chi,\xi) = \chi^{-3}
\xi^3+\xi$. The second option is given by the choice $u=0$ and $v=-2$,
i.e.~$F(\chi,\xi) = \chi^{-3}+\xi^{-2}$. This last model represents a more
traditional point of view where the curvature and matter are independent
parts of the action (see Table~\ref{table01}).

Note that one can also introduce non-locality in such a way that the
coupling parameter \( \alpha \) is a function of the mass  function \(M\)\footnote{As
mentioned in Section~\ref{introduction}, for systems with high degree of
symmetry this mass can be taken as the Bondi-Wheeler mass of the system:

\begin{displaymath}
  M = 4 \pi \int_0^r{ r^2 \rho} \, \text{d}r,
\end{displaymath}

\noindent which for a FLRW space-time turns out to be the Hubble mass with
the choice \( r = ct \)~\citep{carranza13}.}, which from
now on we will assume to be a power law of the form \( M^q \) and so from \eqref{frgeneral} the
function \( F \propto M^q R^p\lmatt^u + \lmatt^v \).
In general terms, the mass \( M \) can be thought of as the causal
mass around any given point in space (for example, the mass within a
Hubble radius for a fundamental observer in cosmological applications).
For this particular situation and following the same procedure as with
the local approach and setting for simplicity \( v = 1 \) it follows
from the trace of the field equations that \( R^{p-1} \propto M^{1-u-q}
/ r^{1-3u} \) and so:

\begin{equation}
   a \propto \frac{ M^{(1-u-q)/(p-1)} }{ r^{(2-p-3q)/(p-1) }}.
	\label{nonlocalac}
\end{equation}

  In this case, the choice \( p = 6q - 3 \) and \( u = 3 - 4q \) yields
MOND for any arbitrary value of \( q \).  The choice \( u=0 \), i.e. \(
q = 3/4 \), yields \( p = 3/2 \), which corresponds to the first metric
extension of MOND built by \citet{bernal11}. All these results are
summarised in Table~\ref{table01}. Due to the complicated nature of the
function \( M \) in the non-local proposal and since it is tradition to
work with locality in physical constructions of gravity, in what follows
we assume locality.

  It is very important to briefly discuss the 
vacuum \( \rho=0 \) consequences of the solutions just found.  In
this case, the function \( F(\chi,\xi) \) of equation~\eqref{frgeneral} is \( F(\chi,\xi) = 0 \) for the
strong curvature-matter coupling and \( F(\chi,\xi) = f( \chi ) 
= \chi^p \)   for the weak curvature-matter coupling.  As such, the 
strong coupling case has validity for non-empty dust models.  The 
weak coupling case in vacuum is not valid since  the right-hand side of equation~\eqref{aprox} diverges for the case \( v = -2 \).  As
such, the local and non-local models we found above are only valid when \( \rho \neq 0 \), except for one non-local exception which occurs for a weak coupling case when \( F(\xi,\chi) = f(R) \propto
R^n \).  As discussed above, this corresponds to 
the non-local  model discussed by~\citet{bernal11} which 
is valid in vacuum and for any value of the energy-momentum tensor.  
Since most of the MONDian astrophysical applications 
require non-vacuum solutions, the above models are quite relevant 
in those scenarios.  Furthermore, for the cosmological applications
explored in this article a non-vacuum dust late-time cosmological 
model is assumed.

\begin{table}
  \begin{center}
    \begin{tabular}{l|l}
      \hline \hline
      \textbf{Local Lagrangian:} \( \mathcal{L}  \propto R^p \lmatt^u + \lmatt^v
        \) \\
      Deep MOND regime obtained when \\
      \( p = -3\) and \( v = u - 2 \) \\ 
      \hline
      (i) Barrientos \& Mendoza (2018) &  $ p= -3 $ \\
      (``weak'' curvature-matter coupling) &  
      $ u=3 $ \\
      $ \mathcal{L} \propto R^{-3} \lmatt^3+ \lmatt$ & $ v=1 $ \\
      \hline
      (ii) Barrientos, Bernal \& Mendoza (this article) &  $ p= -3 $ \\
      (``strong'' curvature-matter coupling) & $ u=0 $ \\
      $ \mathcal{L} \propto R^{-3}+\lmatt^{-2}$ &  
      $ v=-2 $  \\
      \hline \hline
      \textbf{Non-local Lagrangian:} 
        \( \mathcal{L}  \propto M^q R^p \lmatt^u + \lmatt \)  \\ 
      Deep MOND regime obtained when \\
      \( p = 6q - 3 \) and \( u = 3 - 4q \) \\ 
      \hline
      (iii) Bernal, Capozziello, & $u=0$  \\
      Hidalgo \& Mendoza (2011) & $p=3/2$
       \\
      $ \mathcal{L} \propto M^{3/4} R^{3/2} + \lmatt $  
        & $q=3/4 $  \\
      \hline
    \end{tabular}
    \caption[Local and non-local Lagrangians]{\label{table01} \small{The Table shows local and
  non-local Lagrangians \( \mathcal{L} \) in the pure metric approach (but see
  \citet{barrientos16} and \citet{barrientos17} for proposals with
  metric-affine and metric with torsion) which converge to the deep MOND
  regime at the weak-field limit of approximation.  The function \( M \)
  can be thought of as the causal mass (Bondi-Wheeler mass) of the system,
  but it is hard to compute for systems which do not possess a high degree
  of symmetry.  In both, local and non-local approaches it turns out that
  there are an infinite number of actions that converge to the deep MOND
  regime in the weak-field limit of approximation.  For simplicity we
  choose to work in this article with cases (i) and (ii) and as explained 
  in the text, we work
  with local theories of gravity following the traditional approach to
  gravity and so, case (iii) will not be further explored here
  (but see e.g. \citet{mendoza15} and references therein for different
  applications of this proposal).}}
  \end{center}
\end{table}

\section{Cosmography}
\label{cosmography}

In order to constrain  the cosmological
observables in a model-independent way, one can appeal to cosmography, i.e.
using Taylor expansions of  the scale
factor \( a(t) \) with respect to the cosmic time \( t \)  in order to
have a distance-redshift relation that only depends on the FLRW metric
\citep{weinberg}. In this way, the Taylor expansion is given by

\begin{eqnarray}
	H &=& \frac{1}{a} \frac{\text{d}a}{\text{d}t} , \\
	q &=& - \frac{1}{a} \frac{\text{d}^2 a}{\text{d}t^2} H^{-2} , \label{q-param} \\
	j &=& \frac{1}{a} \frac{\text{d}^3 a}{\text{d}t^3} H^{-3} , \label{j-param} \\
	s &=& \frac{1}{a} \frac{\text{d}^4 a}{\text{d}t^4} H^{-4} , \\
	l &=& \frac{1}{a} \frac{\text{d}^5 a}{\text{d}t^5} H^{-5} ,
\end{eqnarray}

\noindent which are called the Hubble, deceleration, jerk, snap and lerk
parameters, respectively. Their values at the present epoch \( t_0 \)
are denoted by a subscript \( 0 \). All models can be characterised by
these parameters: when $ H_0>0 $ produces an expanding universe
and $H_0<0$ a contracting one; $q_0<0$ gives an accelerating expansion,
$q_0>0$ a deceleration one and $q=0$ yields zero acceleration; $H_0=0$
and $q_0=0$ represent a static universe.  If the Taylor expansion is
truncated to lower orders significant deviations for $z\gtrsim 1$ appear.

We use the standard definition of the density parameter $\Omega$:

\begin{equation}
    \Omega:=\frac{8\pi G \rho}{3 H^2},
    \label{omega}
\end{equation}

\noindent such that in standard cosmology $\Omega = \Omega_\text{m} + \Omega_\Lambda + \Omega_\kappa$, with $\Omega_\text{m}$ the matter density parameter, $\Omega_\Lambda$ the effective mass density attributed to dark energy, and $\Omega_\kappa$ the curvature density parameter. In order to explain the accelerated expansion of the Universe without dark energy, we assume $\Omega_\Lambda=0$, and reproduce the SN~Ia observations with the extra terms coming from the generalised gravitational action \eqref{matter-curvature} with curvature-matter couplings. Moreover, the relativistic extensions we are dealing with converge to a MOND-like behaviour in the weak-field limit of the theory. As explained in Section \ref{generalities}, we recover MONDian accelerations that explain the astrophysical observations for systems at accelerations $a < a_0$, without the inclusion of dark matter. In this case, we assume that we are dealing with $\Omega_\text{m}=\Omega_\text{bar}$, the baryonic contribution to the matter density parameter only.

For a universe with $\Omega_\kappa \neq 0$, the
cosmographic parameters are given by~\citep{Kun}:

\begin{eqnarray}
	q_0 &=& \frac{3}{2} \Omega_{\text{m}_0} +\Omega_{\kappa_0}- 1 , \\
	j_0 &=& 1 - \Omega_{\kappa_0}, \\
	s_0 &=& 1 - \frac{9}{2} \Omega_{\text{m}_0} + \Omega_{\kappa_0}^2-\Omega_{\kappa_0}\left(2-\frac{3}{2} \Omega_{\text{m}_0}\right), \\
	l_0 &=& 1 + 3 \Omega_{\text{m}_0} + \frac{27}{2} \Omega^2_{\text{m}_0}+ \Omega_{\kappa_0}^2-\Omega_{\kappa_0}(2-9\Omega_{\text{m}_0}).
\end{eqnarray}

\noindent For the standard $\Lambda$CDM model ($\Omega_\kappa = 0$) the cosmographic parameters reduce to:

\begin{eqnarray}
	q_0 &=& \frac{3}{2} \Omega_{\text{m}_0} - 1 , \\
	j_0 &=& 1 , \\
	s_0 &=& 1 - \frac{9}{2} \Omega_{\text{m}_0} , \\
	l_0 &=& 1 + 3 \Omega_{\text{m}_0} + \frac{27}{2} \Omega^2_{m_0}.
\end{eqnarray}

It is possible to use diverse estimations of the free parameter
$\Omega_{\text{m}_0}$ to obtain the values of the cosmographic parameters. From
the last results of Planck experiment based in a $\Lambda$CDM cosmology
we have: $H_0=(67.4 \pm 0.5)\ \text{km s}^{-1} \text{Mpc}^{-1}$, $\Omega_{\text{m}_0} = 0.315 \pm 0.007$ and $\Omega_{\kappa_0} = 0.0007
\pm 0.0019$ \citep{planck}. Thus, observations are in agreement with
a spatially flat universe. With these results, the predictions for the
$\Lambda$CDM model are: $q_0 = -0.527 \pm 0.0105$, $j_0 = 1$, $s_0= -0.417
\pm 0.0315$ and $l_0 =3.284 \pm 0.0631$.

\section{Cosmology}
\label{cosmo}

  There are many applications of extended theories of gravity to 
cosmology \citep[see e.g.][and references therein]{nojiri-odintsov-review}, but very few of them introduce 
curvature-matter couplings in the gravitational action \cite[cf.][]{harko-lobo-book} and in any case they are only used to
deal with no dark energy.  We are interested in curvature-matter 
couplings with no dark matter, nor dark energy for which the weak-field limit of approximation converges to MOND deep regime.  To do so, we proceed in the following manner.

For the cosmological applications we are interested in, we use a FLRW
metric given by:

\begin{equation}
	\text{d}s^2 = c^2 \text{d}t^2 - a^2(t) \left[ \frac{\text{d}r^2}{1 - \kappa r^2} + r^2 \text{d}\Omega^2 \right],
	\label{FLRW}
\end{equation}

\noindent where $\kappa$ is the curvature of the universe and \( \mathrm{d} \Omega^2  = \mathrm{d} \theta^2 + 
\sin^2 \theta \mathrm{d}\varphi^2 \) is the angular displacement, with \( \theta \) and
\( \varphi \) the polar and azimuthal angles, respectively. From hereafter we adopt the convention: $x^0=ct$ and
$g_{00}=1$.

  Using equations~\eqref{Tpolvo} and~\eqref{lmatt},
the $00$ component of $T_{\mu\nu}^\text{curv}$ (equation \eqref{Tcurv}) takes the value:

\begin{equation}
	T_{00}^\text{curv}=\frac{1}{2}(F-\chi F_\chi)-\frac{3\alpha}{c^2}H\dot{F_\chi},
	\label{T00}
\end{equation}

\noindent where $\dot{\{\}}$ means derivative with respect to
the time coordinate $t$ and $H$ is the Hubble parameter defined
as $H:=\dot{a}/a$.\footnote{In order to obtain
expression~\eqref{T00} the following identity was used: $\Delta
\phi=\partial_\mu \left( \sqrt{-\textsl{g}}\,g^{\mu\nu} \partial_\nu\phi
\right)/\sqrt{-\textsl{g}}$.}

In order to compute the $00$-field equation from~\eqref{field.eqs.eins}, Ricci's 
scalar $R$ and the $00$-component of Ricci's tensor $R_{\mu\nu}$ for the FLRW metric~\eqref{FLRW} are required. Such quantities are given by

\begin{eqnarray}
    R &=& -6\left(\frac{a \ddot{a}+{\dot{a}}^2+ \kappa c^2}{c^2 a^2}\right) ,
	\label{Ricci} \\
	R_{00} &=& -\frac{3\ddot{a}}{c^2 a} .
	\label{Ricci00}
\end{eqnarray}

\noindent Therefore, from the $00$-component of equations~\eqref{field.eqs.eins} the Hubble parameter is
\begin{equation}
	H^2=\frac{c^2}{3\alpha F_\chi}\left[
	\frac{\rho c^2 F_\xi}{2\lambda}+\frac{F-\chi F_\chi-\xi F_\xi }{2}-\frac{3H\alpha}{c^2}\dot{F_\chi}\right]-\frac{\kappa c^2}{a^2}.
	\label{00}
\end{equation}

Notice that for the spatial field equations required, due to the symmetry, we have only one independent equation, so we choose $i=j=1$. In order to obtain the expression for this component, $T_{11}^\text{curv}$ and
$R_{11}$ are needed. Such tensor components are given by

\begin{equation}
    T_{11}^\text{curv}=\frac{a^2}{1 - \kappa r^2}\left[-\frac{1}{2}(F-\chi F_\chi)+\frac{\alpha}{c^2}(\ddot{F_\chi}+2H\dot{F_\chi})\right],
    \label{11Tcurv}
\end{equation}

\noindent and

\begin{equation}
    R_{11} = \frac{2 \kappa c^2+a\ddot{a}+2\dot{a}^2}{c^2(1 - \kappa r^2)}.
    \label{R11}
\end{equation}

\noindent Substituting these relations into
equation~\eqref{field.eqs.eins}, the following equation for the \( 11 \) 
radial-component of the field equations is obtained:

\begin{eqnarray}
    -\frac{\kappa c^2+2a\ddot{a}+\dot{a}^2}{a^2c^2}&=&\frac{1}{\alpha F_\chi}\left[\frac{PF_\xi}{2\lambda}-\frac{\left(F- \chi F_\chi - \xi F_\xi \right)}{2}\right. \nonumber \\
    & &+\frac{\alpha}{c^2}\left(\ddot{F_\chi}+2H\dot{F_\chi}\right)\bigg].
    \label{11field}
\end{eqnarray}

\subsection{``Weak'' curvature-matter coupling model}
\label{subsec-addition}

  Let us assume the following form for the function $F(\chi,\xi)$: 

\begin{equation}
	F(\chi,\xi)=f(\chi)+ g(\xi),
	\label{f}
\end{equation}

\noindent this is, a ``weak'' curvature-matter coupling, as curvature $\chi$ and matter $\xi$ appear independently into the gravitational action and in the resulting field equations. With this, the Hubble parameter~\eqref{00} turns into:

\begin{equation}
	H^2 = \frac{c^2}{3\alpha f_\chi}\left[
	\frac{\rho c^2 g_\xi}{2\lambda}+ \frac{ f+g-\chi f_\chi - \xi g_\xi }{2} -\frac{3H\alpha}{c^2}\dot{f_\chi}\right] -\frac{\kappa c^2}{a^2},
	\label{00.suma}
\end{equation}

\noindent where $f_\chi := \text{d}f/\text{d}\chi$ and $g_\xi := \text{d}g/\text{d}\xi$. By considering a dust universe it 
follows that $T_{00}=\lmatt=\rho c^2$. Thus $\xi=\rho c^2/\lambda$ and so, the previous equation reduces to

\begin{equation}
	H^2=\frac{c^2}{3\alpha f_\chi}\left[\frac{f+g-\chi f_\chi}{2}-\frac{3H\alpha}{c^2}\dot{f_\chi}\right]-\frac{\kappa c^2}{a^2}.
	\label{00.suma.dust}
\end{equation}

\noindent Given the results from \citet{planck} mentioned in the previous
section, from now on we will assume a flat Universe with $\kappa=0$.

 Let us assume a power-law
distribution for the functions $f=\chi^\gamma$ and $g=\xi^\beta$, with
$\gamma$ and $\beta$ real constants, so that

\begin{equation}
    F(\chi,\xi) = \chi^\gamma + \xi^\beta .
    \label{fsuma}
\end{equation}

\noindent With this choice, equation~\eqref{00.suma.dust} becomes

\begin{equation}
	H^2=\frac{c^2}{3\alpha \gamma \chi^{\gamma-1}}\left[\frac{(1-\gamma)\chi^\gamma+
	\xi^\beta}{2}-\frac{3H\alpha\gamma}{c^2}\frac{d}{dt}\chi^{\gamma-1}\right].
	\label{00.dust.power}
\end{equation}

In order to simplify the previous expression, two cosmographic parameters
are particularly useful: the deceleration $q$ and the
jerk $j$ given by equations~\eqref{q-param} and~\eqref{j-param}
respectively. With such definitions, Ricci's
scalar $R$ shown in relation~\eqref{Ricci} and its time derivative $\dot{R}$
for a flat universe take the following form:

\begin{equation}
	R=-\frac{6H^2}{c^2}(1-q) \quad \mbox{and} \quad \dot{R}=RH\left(\frac{j-q-2}{1-q}\right),
	\label{Rs}
\end{equation}

\noindent With all this, it follows that: 

\begin{equation}
	\frac{\text{d}}{\text{d}t}\left(\chi^{\gamma-1}\right)=\chi^{\gamma-1}H
	(\gamma-1)\left(\frac{j-q-2}{1-q}\right).
	\label{tderivative}
\end{equation}

\noindent After substitution of this last result and expressions \eqref{Rs} into 
equation \eqref{00.dust.power}, we obtain the following relation between the Hubble parameter $H$
and the matter density $\rho$:

\begin{equation}
	H^{2\gamma}=\frac{(q-1)^{1-\gamma}c^{2(\gamma+\beta)}}{6^\gamma \gamma \lambda^\beta \alpha^\gamma Z}
	\rho^\beta,
	\label{Hubble}
\end{equation}

\noindent where the function $Z$ is defined as

\begin{equation}
	Z := 1+(\gamma-1)\left(\frac{q-1}{\gamma}+\frac{j-q-2}{1-q}\right).
	\label{Z}
\end{equation}

\noindent After using the definition of the density parameter $\Omega$ (equation \eqref{omega}) and the coupling constants given in \eqref{constantes}, equation~\eqref{Hubble} can be simplified to yield a modified Friedmann equation for the weak curvature-matter
model:

\begin{equation}
	H=\frac{a_0}{c}\left[\frac{(q-1)^{1-\gamma}}{6^\gamma \gamma k^\gamma k'^\beta Z}\left( \frac{3\Omega}{8\pi}\right)^\beta
	\right]^{1/2(\gamma-\beta)}.
	\label{Hubble2}
\end{equation}

By using equations \eqref{q-param}, \eqref{j-param} and \eqref{fsuma}, the 11-component of the field 
equations~\eqref{11field} can be written as:

\begin{eqnarray}
    \frac{H^2}{c^2}(2q-1) &=& \frac{1}{\alpha\gamma\chi^{\gamma-1}}\left\{\frac{1}{2}\left[(\beta-1)\xi^\beta+(\gamma-1)\xi^\gamma\right]\right. \nonumber \\
    & &\left.+\frac{\alpha}{c^2}\left(\ddot{f_\chi}+2H\dot{f_\chi}\right) \right\}.
    \label{11suma.plano}
\end{eqnarray}

In order to compute $\ddot{f_\chi}$, the following expressions are useful:

\begin{eqnarray}
    \dot{H}&=&-H^2(1+q),\\
    \dot{q}&=&-H(j-q-2q^2),\\
    \ddot{H}&=&H^3(2+3q+j),\\
    \dot{j}&=&H(s+2j+3jq),\\
    \dddot{H}&=&-H^4(6+12q+4j+3q^2-s),
    \label{derivadasqj}
\end{eqnarray}

\noindent and so, using the previous relations and
equation~\eqref{tderivative} we obtain:

\begin{equation}
    \ddot{f_\chi}=\frac{\gamma(\gamma-1)}{1-q}\chi^{\gamma-1}H^2A,
    \label{ftt}
\end{equation}

\noindent where:

\begin{eqnarray}
    A &:=& (j-q-2) \left[ (\gamma-1) \left( \frac{j-q-2}{1-q} \right) - 1 - q \right] + s - q \nonumber \\
    && + 3j (1 + q) - 2 q^2 + \frac{(j-q-2)(q+2q^2-j)}{1-q}.
    \label{A}
\end{eqnarray}

 Substitution of this last result with the definitions of $\chi$
and $\xi$, together with equation \eqref{tderivative}, into relation
\eqref{11suma.plano} yield:

\begin{eqnarray}
\label{11suma.plano.2}
    \frac{H^2}{c^2}\left[2q-1+\frac{1-\gamma}{1-q}\left(A+2(j-q-2)\right)+\frac{3}{\gamma}(\gamma-1)(1-q)\right]\nonumber\\
    =\frac{\beta-1}{2\alpha\gamma}\left(\frac{\rho c^2}{\lambda}\right)^\beta\left[\frac{6\alpha H^2}{c^2}(q-1)\right]^{1-\gamma}.
\end{eqnarray}

\noindent In order to simplify this relation, we rewrite equation
\eqref{Hubble} as

\begin{equation}
    \frac{6Z H^2}{c^2}=\frac{1}{\alpha\gamma}\left(\frac{\rho c^2}{\lambda}\right)^\beta \left[\frac{6\alpha H^2}{c^2}(q-1)\right]^{1-\gamma}.
    \label{intermediotmp}
\end{equation}

\noindent Substitution of this relation into \eqref{11suma.plano.2} yields:

\begin{equation}
    2q-1+\frac{1-\gamma}{1-q}A+\frac{3}{\gamma}(\gamma-1)(1-q)+3(1-\beta)Z=0.
    \label{s.qj}
\end{equation}

\noindent The previous equation involves exclusively the cosmographic
parameters $q$, $j$ and $s$.  Thus, we can express the snap parameter
$s$ as a function of the deceleration $q$ and jerk $j$ parameters, i.e.
$s=s(q,j)$, decreasing the number of free parameters to fit with
SNe~Ia observations as explained in Section \ref{ajuste}.

\subsubsection{Geodesic equation}
\label{geo}

The contravariant form of the field equations \eqref{field.eqs.} for the
weak curvature-matter model given in \eqref{f} is:

\begin{eqnarray}
	&&\alpha f_\chi R^{\mu\nu} -\frac{1}{2}g^{\mu\nu}f+\alpha \left(g^{\mu\nu}\Delta-
	\nabla^\mu\nabla^\nu \right)f_\chi\nonumber \\
	&&=\frac{1}{2}g^{\mu\nu}\left(g- \xi g_\xi \right)+\frac{g_\xi T^{\mu\nu}}{2\lambda}.
	\label{geo.field}
\end{eqnarray}

\noindent Taking the covariant divergence of the last equation and bearing in mind that $\nabla_\mu f=f_\chi\partial_\mu \chi$
and $\nabla_\mu g=g_\xi\partial_\mu \xi$, we obtain: 

\begin{align}
	\label{geo.field2}
    &-\frac{1}{2}g^{\mu\nu}\xi\nabla_\mu g_\xi+\frac{\nabla_\mu g_\xi T_{\mu\nu}+g_\xi\nabla_\mu T^{\mu\nu}}{2\lambda} = \\
    & \alpha f_\chi \nabla_\mu R^{\mu\nu} +\alpha R^{\mu\nu}\nabla_\mu f_\chi +\alpha g^{\mu\nu}(\partial_\mu\Delta-
	\Delta\nabla_\mu)f_\chi \nonumber 
    \\ & -\frac{1}{2}g^{\mu\nu}f_\chi \partial_\mu
	\chi. \nonumber
\end{align}

\noindent Since $(\Delta\nabla_\mu - \nabla_\mu\Delta) \phi = R_{\mu\nu}
\nabla^\nu \phi$ \citet{Koivisto, Sotiriou}, the previous equation becomes

\begin{eqnarray}
    \alpha f_\chi \nabla_\mu \left( R^{\mu\nu} - \frac{1}{2} g^{\mu\nu} R\right) &=& \frac{1}{2\lambda} \left(  T^{\mu\nu} \nabla_\mu g_\xi - g_\xi \nabla_\mu T^{\mu\nu} \right) \nonumber \\
    && -\frac{1}{2}g^{\mu\nu}\xi\nabla_\mu g_\xi.
    \label{bianchi}
\end{eqnarray}

 Using Bianchi's identities, or equivalently the  null divergence
of Einstein's tensor, the left-hand side of the previous equation
vanishes. Thus, the conservation equation is given by

\begin{equation}
    \nabla_\mu T^{\mu\nu} = \left( T^{\mu\nu} - g^{\mu\nu} \lmatt \right) \nabla_\mu \mbox{ln} \, g_\xi.
    \label{cons}
\end{equation}

\noindent We are interested in the 00 time-component of the last
expression. Since the energy-momentum and the metric tensors are
diagonal, and $T^{00}=\lmatt=\rho c^2$ for dust, the right-hand side
of equation~\eqref{cons} is zero. Thus, we effectively have a conserved
quantity, which is a desirable characteristic of our theory:

\begin{equation}
    \nabla_\mu T^{\mu 0}=0.
    \label{00geo}
\end{equation}

\noindent With the FLRW metric~\eqref{FLRW}, this conservation equation yields:

\begin{equation}
    \dot{\rho}+3H\rho=0,
    \label{densidad}
\end{equation}

\noindent and so:

\begin{equation}
    \rho(t) = C a^{-3}(t),
    \label{density.dust}
\end{equation}

\noindent where \( C \) is an integration constant.  Note that
equation~\eqref{density.dust} is an expected result for a dust universe
where matter is conserved.

\subsection{``Strong'' curvature-matter coupling model}
\label{subsec-product}

Inspired in the article by \citet{barrientos18}, we explore a ``strong'' curvature-matter coupling for the function $F(\chi,\xi)$:

\begin{equation} 
        F(\chi,\xi)=\chi^\gamma \xi^\beta + \xi.
         \label{fproduct} 
\end{equation} 

\noindent In this case, the matter Lagrangian $\xi$ appears as a linear independent term in addition to the ``strong'' coupling term between curvature $\chi$ and matter $\xi$ as a multiplication. With this function, equation~\eqref{00} for the Hubble parameter
in the dust case and curvature $\kappa=0$ is given by:

\begin{eqnarray} H^2&=&\frac{c^2}{3\alpha \gamma
        \chi^{\gamma-1}\xi^\beta}\left[\frac{8\pi G\alpha}{c^2}\rho
    +\frac{1}{2}(1-\gamma)\chi^\gamma\xi^\beta\right.\nonumber
         \\ & &\left.-\frac{3H\alpha
         \gamma}{c^2}\frac{\text{d}}{\text{d}t} \left(\chi^{\gamma-1}\xi^\beta \right)\right].
   \label{001dust} 
   \end{eqnarray}

In order to manipulate the last expression, the following standard
assumptions are made: 

\begin{equation}
        \rho=\rho_0\left(\frac{a}{a(t_0)}\right)^\tau \quad \mbox{and}
        \quad a=a(t_0)\left(\frac{t}{t_0}\right)^\sigma,
        \label{supa}
\end{equation}

\noindent for real constants $\tau$ and $\sigma$. The value \( \tau =
-3 \) follows from equation~\eqref{density.dust}. By taking the derivative with respect to the time coordinate
we have

\begin{equation}
         \dot{\rho}=\tau\rho H \quad \mbox{and} \quad \dot{H}=-\frac{H^2}{\sigma}.
         \label{utiles}
\end{equation}

 Following the same procedure as the one used in
Subsection~\ref{subsec-addition}, we use equations~\eqref{Rs} for $R$
and $\dot{R}$ and equations~\eqref{utiles} to obtain the time derivative
of $\chi$:

\begin{equation}
        \frac{\text{d}}{\text{d}t} \left(\chi^{\gamma-1}\xi^\beta\right)
      = \chi^{\gamma-1}\xi^\beta H\left[(\gamma-1)
      \left(\frac{j-q-2}{1-q}\right)+\beta\tau\right].
\label{tderivative2}
\end{equation}

\noindent Substitution of this last result and expressions \eqref{Rs} into
equation~\eqref{001dust} yields a relation between the
Hubble parameter and the matter density:

\begin{equation}
        H^{2\gamma} = \left[6(q-1)\right]^{1-\gamma}\frac{8\pi G}{3\gamma Z'}
        \frac{\lambda^\beta\alpha^{1-\gamma}}{c^{2(1-\gamma+\beta)}}
        \rho^{1-\beta},
        \label{Hubble.prod}
\end{equation}

\noindent where
         
\begin{equation}
        Z' := 1+(1-\gamma)\left[\frac{1-q}{\gamma}-\frac{j-q-2}{1-q}\right]+\tau\beta.
        \label{Z2}
\end{equation}

With the values for the coupling constants $\alpha$ and $\lambda$
given in \eqref{constantes}, together with the density
parameter $\Omega$ defined in \eqref{omega}, the Hubble
parameter~\eqref{Hubble.prod} reduces to:

\begin{equation}
	H=\frac{a_0}{c}\left[ \left[6(q-1)\right]^{1-\gamma} \frac{8\pi k^{1-\gamma} k'^\beta}{3\gamma Z'}\left( \frac{3\Omega}{8\pi}\right)^{1-\beta}
	\right]^{1/2(\gamma+\beta-1)}.
	\label{Hubble2.prod}
\end{equation}

\noindent With this function, the space 11-component of the field
equations~\eqref{11field} can be written as:

\begin{eqnarray}
    \frac{H^2}{c^2}(2q-1)&=&\frac{1}{\alpha\gamma\chi^{\gamma-1}\xi^\beta}\left[\frac{1}{2}(\beta+\gamma-1)\chi^\gamma\xi^\beta\right. \nonumber \\
    & &\left.+\frac{\alpha}{c^2}\left(\ddot{f_\chi}+2H\dot{f_\chi}\right) \right],
    \label{11suma.prod}
\end{eqnarray}

\noindent where $\dot{f_\chi}$ is given by equation~\eqref{tderivative2}. 
The second time derivative $\ddot{f_\chi}$ is given by:

\begin{equation}
    \frac{\text{d}^2}{\text{d}t^2} \left(\chi^{\gamma-1}\xi^\beta\right)=\chi^{\gamma-1}\xi^\beta H^2B,
    \label{ftt2}
\end{equation}

\noindent where:

\begin{eqnarray}
    B &:=& \left[(\gamma-1)\left(\frac{j-q-2}{1-q}\right)+\beta\tau\right]\left[(\gamma-1)\left(\frac{j-q-2}{1-q}\right)\right.\nonumber \\
    & &\left.+\beta\tau-1-q\bigg]\right.+\left(\frac{\gamma-1}{1-q}\right)\bigg[\left.s+3j+3jq-q-2q^2\right.\nonumber \\
    & &\left.+\frac{(j-q-2)(q+2q^2-j)}{1-q}\right].
    \label{B}
\end{eqnarray}

Substitution of equations~\eqref{ftt2} and~\eqref{tderivative2}
into~\eqref{11suma.prod} yields:

\begin{eqnarray}
    \frac{3}{\gamma}(\beta+\gamma-1)(q-1)+1-2q+B& &\nonumber \\
    +2\left[(\gamma-1)\left(\frac{j-q-2}{1-q}\right)+\beta\tau\right]&=&0.
    \label{s.qj2}
\end{eqnarray}

\noindent The last expression is the functional $s=s(q,j)$ equation
for the strong curvature-matter coupling model. Unlike its analogous for the addition model in
equation \eqref{s.qj}, note that in order to obtain relation \eqref{s.qj2}
the time 00-component of the field equations was not required.

\section{Type Ia Supernovae best fit}
\label{ajuste}

By precise mappings of the distance-redshift relation to redshifts $z
\lesssim 1.3$, type Ia supernovae (SNe~Ia) remain one of the most robust
probes for the accelerating expansion of the Universe at late times. In
order to cosmologically test the models presented in Section \ref{cosmo},
we used the distance-redshift data from the Supernova Cosmology Project
(SCP) Union 2.1 \citep{SCP-Union2012}.

Assuming that SNe~Ia events form a homogeneous class for distance
estimation, having on average the same intrinsic luminosity for all
redshifts, a reasonably linear model for the distance modulus to construct
the Hubble diagram is given by the following relation \citep{peebles}\footnote{In fact,
it is shown in equation~(13.52) of \citet{peebles} that the distance modulus is given by:

\begin{displaymath}
  \mu(z) =  25 + 5 \log\left[ 3000 d_L(z) \right] - 5 \log h(z),
\end{displaymath}

\noindent where the distance modulus \( d_{L}(z) = (1 + z) r(z)  \).  
Since \( 5 \log(3000) = 17.3856 \) then equation~\eqref{modulus} follows directly.
}:

\begin{equation}
    \mu(z) = 5 \log \left[\frac{H_0 d_L(z)}{c}\right] - 5 \log h(z) + 42.3856,
    \label{modulus}
\end{equation}

\noindent where $H_0$ is the Hubble constant at the present epoch, $h$ is the normalised Hubble constant defined by $h := H_0 / (100 \, \text{km/s} \cdot \text{Mpc})$, 
and $d_L(z)$ is the luminosity distance \citep{Visser} given by:

\begin{eqnarray}
    d_L(z) = \frac{c}{H_0} \left[z+\frac{1}{2}(1-q_0)z^2
    -\frac{1}{6} \left(1-q_0-3q_0^2+j_0 \right)z^3\right. \nonumber \\
    \left. +\frac{1}{24} \left(2-2q_0-15q_0^2-15q_0^3+5j_0+10q_0 j_0+s_0\right)z^4 + ...\right], \nonumber \\
    \label{distance}
\end{eqnarray}

\noindent for a flat universe with curvature $\kappa=0$, and $H_0, \ q_0, \ j_0$ and $s_0$ the cosmographic
parameters evaluated at the present epoch. It is important to mention that
equations \eqref{modulus} and \eqref{distance} are entirely
model-independent. Therefore, with the observational data for the
redshift $z$ and the modulus distance $\mu$ from SCP Union 2.1 data,
we test the models \eqref{fsuma} and \eqref{fproduct} to callibrate their
corresponding free parameters.

To do so, we substitute equation \eqref{distance} into \eqref{modulus}.
In this way the distance modulus turns into a function of $z$, $H_0$,
$q_0$, $j_0$ and $s_0$. Thus, initially the expression has four
free parameters since $z$ is given by the observations.  From the
00-component of Friedmann's equations for each proposal, we have a
relation for $H_0$ as a function of $\Omega_0$, $q_0$ and $j_0$ (see
equations~\eqref{Hubble2} and \eqref{Hubble2.prod}). Note that we still have four
free parameters since  we have only changed $H_0$ for $\Omega_0$. But from the
space 11-component of Friedmann's equations (see \eqref{s.qj} and \eqref{s.qj2})
we can obtain an expression for $s_0$ as a function of $q_0$ and $j_0$;
in this manner the number of unknown free parameters is diminished to three,
$q_0$, $j_0$ and $\Omega_0$.

In order to compute the values for the free parameters, a fit was performed using 
the free software \textit{gnuplot} (\url{www.gnuplot.info}). 
First, the value for the constants 
$c$ and $a_0$ are given. Since the interest lies in the functions $ F(\chi,\xi) = \chi^{-3}
\xi^3+\xi$ and $F(\chi,\xi) = \chi^{-3}+\xi^{-2}$, the set of values: $\gamma=-3$, $\beta=3$ and 
$\gamma=-3$, $\beta=-2$ are introduced for the corresponding model. According to equations \eqref{Hubble2} 
and \eqref{Hubble2.prod}, the coupling constants $k^3k'^3$ and $k^4k'^3$ must be given for each model too. With all this,
the functions $Z_0=Z_0(q_0,j_0)$, $H_0=H_0(q_0,j_0,\Omega_0)$, $s_0=s_0(q_0,j_0)$, 
$d_L=d_L(z,q_0,j_0)$ and $\mu=\mu(z,q_0,j_0,\Omega_0)$ can now 
be computed. Initial values 
for $q_0$, $j_0$ and $\Omega_0$ must be specified for every fit. 
We used the fit function command in \textit{gnuplot} for the calibration of the
cosmographic parameters \(q\), \(j\) and the density parameter
\( \Omega_0 \) given in our model.  This function uses non-linear
and linear least squares methods.



  With the previous information \textit{gnuplot's} fit function
returns the information of the number of iterations employed for 
the converged fit, the correlation matrix between
the parameters, the best fit value for the parameters and its asymptotic standard error, the final sum of the squares 
of residuals (SSR), the root mean squares of residuals and the p-value for the $\chi$-square distribution.

\section{Results}

\subsection{``Weak'' curvature-matter coupling model}
\label{weak-curvature}

The best fit results from the SNe~Ia observations for the model $F(\chi,\xi)=\chi^{-3}+\xi^{-2}$ for the 
cosmographic parameters and the correlation matrix are shown in Table~\ref{table02}. The set of
initial values for this fit was: $q_0=-0.8$, $j_0=0.5$ and $\Omega_0=0.8$. The relation between 
$k$ and $k'$ is given in equation \eqref{ks}.

\begin{table}
	\begin{center}
		\begin{tabular}{|c|c|}
		\hline
		\multicolumn{2}{|c|}{$F(\chi,\xi)=\chi^{-3}+\xi^{-2}$} \\ 
		\hline
		$q_0$ & $-0.428081\pm 0.05646$ \\
		$j_0$ & $-0.345827\pm 0.08342 $ \\
		$\Omega_0$ & $1.85257 \pm 0.28900$ \\
		\hline
		\end{tabular}
        \begin{tabular}{|c|c|c|c|}		
		 \hline
         & $q_0$ & $j_0$ & $\Omega_0$\\ 
         \hline
         $q_0$ & 1.000 & & \\
         $j_0$ &-0.846 & 1.000 & \\
         $\Omega_0$ & -0.997 & 0.886 & 1.000\\
         \hline
		\end{tabular}
	\end{center}
\caption[Table02]{\small{Left: Best fit results
	for the``weak'' curvature-matter coupling model (see Subsection~\ref{subsec-addition}), which
	corresponds to case (ii) of Table~\ref{table01}.  The 
	deceleration $q_0$, jerk $j_0$ and density parameter $\Omega_0$ are shown with their corresponding errors. Right: Correlation matrix for the best 
    fit values reported. The SSR for this model is: \( 
    563.621 \).
    }}
\label{table02}
\end{table}

\begin{figure}
  \includegraphics[scale=0.7]{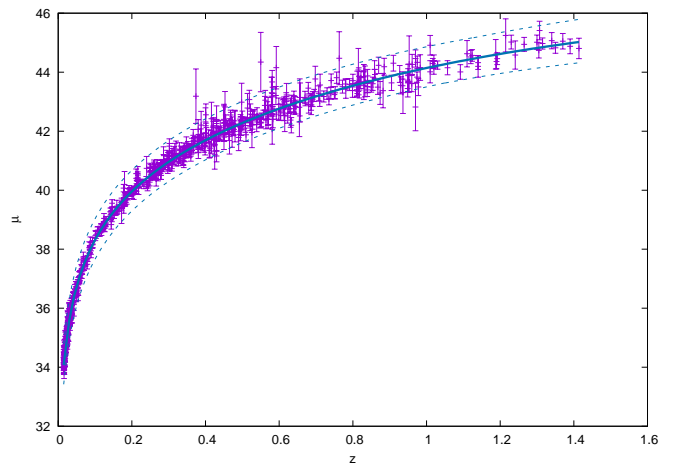}
  \caption[Figure01]{\small{Apparent magnitude \( \mu \) vs.~redshift \(
z \) Hubble diagram from the Union 2.1 SNe~Ia data (dots with their
corresponding error bars) and the best fit from our addition model (case
(i) in Table~\ref{table01}): $F(\chi,\xi)=\chi^{-3}+\xi^{-2}$. The solid
line represents the distance modulus $\mu(z)$ from the best fit to the
data of the model; the dashed lines represent the maximum and minimum
errors of the fit.}}
	\label{figsuma}
\end{figure}

A graph of $\mu$ vs.~$z$ for the observational data is shown in Figure~\eqref{figsuma}.

 Notice that from the best fit $q_0 < 1$, meaning
that the Universe is accelerating. The resulting Hubble constant $H_0$
(equation~\eqref{Hubble2}) obtained from substitution of the values
reported in Table~\ref{table02}  is:

\begin{equation}
    H_0=69.511934^{+23.495497}_{-18.893123} \text{ km s}^{-1} \text{Mpc}^{-1} ,
    \label{valorH0suma}
\end{equation}

\noindent which is a very good value, but with large errors. 

\subsection{``Strong'' curvature-matter coupling model}
\label{strong-curvature}

By performing an analogous procedure as the one in subsection~\ref{weak-curvature} for the ``strong'' curvature-matter coupling model
$F(\chi,\xi)=\chi^{-3}\xi^3 + \xi$, the values of the cosmographic parameters
for this function are displayed in Table~\ref{table03}, with their corresponding correlation matrix
of the fit. The initial values for the parameters in this fit were: $q_0=-0.3$, $j_0=0.5$ and $\Omega_0=1.0$.
The relation between $k$ and $k'$ is the following: $k^4k'^3  = 9 / 4^5 \pi^2\approx 8.9\times 10 ^{-4}$  \citep[see the Erratum at the end of the article by][in \url{https://arxiv.org/abs/1808.01386}]{barrientos18}.

\begin{table}
	\begin{center}
		\begin{tabular}{|c|c|}
		\hline
		\multicolumn{2}{|c|}{$F(\chi,\xi)=\chi^{-3}\xi^{3} + \xi $} \\ 
		\hline
		$q_0$ & $-0.417946\pm 0.06891$ \\
		$j_0$ & $-0.240291\pm 0.1493 $ \\
		$\Omega_0$ & $79.0745 \pm 1.632 $ \\
		\hline
		\end{tabular}
		 \begin{tabular}{|c|c|c|c|}
         \hline
         & $q_0$ & $j_0$ & $\Omega_0$\\ 
         \hline
         $q_0$ & 1.000 & & \\
         $j_0$ &-0.914 & 1.000 & \\
         $\Omega_0$ & -0.244 & -0.164 & 1.000\\
         \hline
    \end{tabular}
	\end{center}
		\caption[Table03]{\small{Left: Best-fit results for the ``strong'' curvature-matter coupling model
		(see Subsection~\ref{subsec-product}) corresponding to case
		(i) of Table~\ref{table01}. The deceleration $q_0$,
		jerk $j_0$ and density parameter $\Omega_0$ are shown with their corresponding errors. Right: Correlation matrix for the best-fit values reported. 
		The SSR for this model is: \(  639.521 \).
		}}
\label{table03}
\end{table}

\begin{figure}
  \includegraphics[scale=0.7]{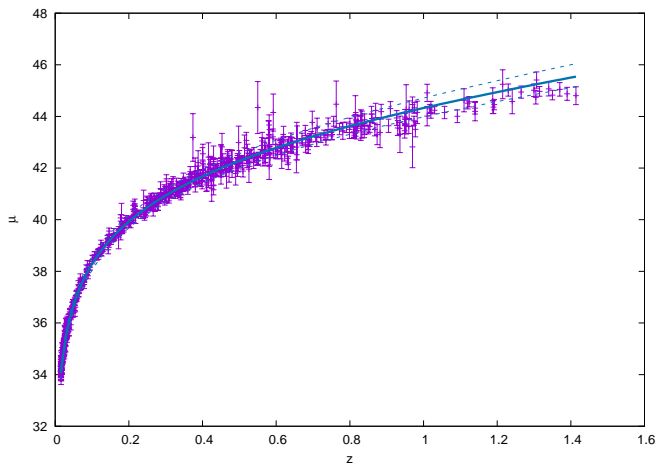}
  \caption[Figure03]{\small{Distance modulus $\mu(z)$ from the Union 2.1 SNe~Ia
  data (dots with their corresponding error bars) and the best fit from
  the product model $F(\chi,\xi)=\chi^{-3}\xi^3 + \xi$. The solid line is the
  best fit to the observations; the dashed lines represent the errors
  in the fit.}}
	\label{figprod}
\end{figure}

  Figure \ref{figprod} shows the Hubble diagram for this model. 

For this model, the value of the deceleration parameter $q_0$ is similar to the 
"weak" coupling model. Therefore, both models describes an accelerating Universe.

The resulting value for the Hubble parameter within this proposal is
given by: 

\begin{equation}
    H_0=70.363944^{+5.958497}_{-5.550341} \text{ km s}^{-1} \text{Mpc}^{-1} .
    \label{valorH0prod}
\end{equation}

\section{Discussion}
\label{discussion}

  As mentioned in Section \ref{introduction}, the construction of a relativistic theory of MOND, or more precisely a relativistic theory
that converges in the weak field limit to the Tully-Fisher law, has so far been a complicated issue.   
Motivated by general relativity, it has always been desirable to build a pure metric relativistic theory of MOND with no extra
(scalar, vector or tensor) fields.  In this article we have shown how this is possible in an 
infinite number of ways by allowing curvature-matter couplings in the gravitational action.  
Specifically, we showed that it is possible to have an infinite number of 
local and non-local Lagrangians with  curvature-matter couplings for which their weak field
limit of approximation reproduce the Tully-Fisher law (deep MOND regime) and it is coherent with
the deflection of massless (light) particles observed in individual, groups and clusters of 
galaxies.  These general results were presented in Table~\ref{table01} and for the cosmological 
applications presented in this article, we decided to work only with a local theory of gravity
in which two simplest models were studied: (a) a weak curvature-matter coupling and (b) a
strong curvature-matter coupling. Of course it would be possible to use any non-local 
model for cosmological applications \citep[see e.g. the work by][where the non-local model example presented in Table~\ref{table01} was applied to the accelerated expansion of the Universe fitting SN~Ia observations]{carranza13}, but since it is tradition in gravitational studies to work
with locality we decided to follow that path.

It is important to note that in \citet{barrientos16, barrientos17} the feasibility of having the connection $\boldsymbol{\Gamma}$,
both symmetric (Palatini) and non-symmetric (torsion), as an independent field responsible for the MONDian behaviour of the
gravitational phenomenon in the correct scale was proposed, 
Despite the positive results found in those works, the cosmological consequences of this approach were not explored in this article 
because of the following facts. In studies of the  pure metric formalism, the term
responsible for the MONDian behaviour is that which contains derivatives of the Ricci scalar
(the fourth-order derivative term of the metric). This term does not appear
in the Palatini metric-affine formalism. In this approach that term comes from the conformal relation between the Levi-Civita metric and the general one. This fact makes the resultant field equations in the Jordan frame very complex. Also, an
extra linear assumption about the transformation from the Jordan frame to the Einstein one is necessary to achieve the desirable
expression in the weak field limit. On the other hand, in the torsion formalism the field equations only apply for the symmetric
part of the Ricci's tensor. The disadvantages mentioned previously for the Palatini metric-affine formalism also appear in the
torsion approach.  In order to avoid this, derivatives of the matter lagrangian were introduced in the matter sector of the action. 
The meaning n of these derivatives is not clear and yield an unwanted complicated field equations. Therefore, the study of  matter-curvature coupling theories can be safely  restricted  to the pure metric formalism.

  The conception of dark energy as responsible for the accelerated expansion of the Universe gained momentum 
among other ideas because its success in fitting type Ia supernovae observations. In \citet{Perlmutter99} 
an analysis between 46 type Ia supernovae and  general relativity through a Friedmann model with dark energy was performed. In that work, the 
cosmographic parameters $q_0$, $j_0$ and $s_0$ were written as functions of the matter density parameter $\Omega_m$ and the dark energy density parameter $\Omega_\Lambda$ (see e.g. Section~\ref{cosmography}). Using  
equation~\eqref{distance}, a theoretical function for the distance 
modulus $\mu(z)$ is then
built where $\Omega_m$ and $\Omega_\Lambda$ are the only two free parameters of the theory. It was found that in order to reproduce
the observed modulus distance of those 46 type Ia supernovae, the values for the density parameters must have the following values:
$\Omega_m\approx 0.3$ and $\Omega_\Lambda\approx 0.7$. In the work presented in this article we made a similar analysis, using a single
matter density parameter \( \Omega_m \).  Its particular functional form as function of the cosmographic parameters is cumbersome but can directly be obtained form equation~\eqref{Hubble2} and~\eqref{Hubble2.prod} for each 
model. 
s. In any case, as mentioned in Section~\ref{ajuste}, it is possible to obtain all cosmographic parameters by fitting only three of them.
Our choice was to fit 
$ \omega_m $, $q_0$ and $j_0$.  As shown in Section~\ref{ajuste}, the found values for the cosmographic
parameters of our two models are in good agreement with the values
expected from the cosmological concordance model \citep{planck}.  
Therefore, the weak and strong curvature-matter coupling models studied in this work are feasible candidates to explain the accelerated expansion of the Universe
without the introduction of any dark component.

  It is very important to mention once more that the matter density
parameter 
\( \Omega \) defined in equation~\eqref{omega} correspond to the 
standard definition used in cosmological studies using general
relativity.  Furthermore, it follows that in standard cosmology
the sum of all density parameters (baryonic plus dark matter, 
dark energy, curvature and radiation) is equal to one when evaluated at the present epoch \citep{longair-galaxy-formation}.  
This result is a direct consequence of the standard density 
parameters definitions directly motivated by the standard Friedmann
equation.  The use of a more general gravitational action with
curvature-matter couplings in this article yields a more extended
gravitational field equations~\eqref{fieldeqs} and in consequence
the Firedmann equations~\eqref{Hubble2} and~\eqref{Hubble2.prod}
obtained for our two particular strong and weak models greatly 
differ from the standard Friedmann equation of general relativity.
This means that in order to keep the definition of e.g. the matter
density parameter we would have to do it as it was done in standard
cosmology: it will be the ratio of the matter density to the 
critical density that closes the universe in the absence of any other density parameters.  For simplicity and in the spirit of 
avoiding confusion, we 
preferred to keep the standard definition of the density parameter
with the immediate consequence that \( \Omega_m \) will not be 
one for a flat dust universe without dark components.  Furthermore,
the \(\Omega_m \) value will also be model dependent as it can be 
seen from the results reported in Section~\ref{ajuste}.




  Equations~\eqref{Hubble2} and~\eqref{Hubble2.prod} can be
written as:

\begin{equation}
    H=\frac{a_0}{c}\Omega_\text{eff},
    \label{effective}
\end{equation}

\noindent where the definition of $\Omega_\text{eff}$ is model dependent but its numerical value
should not. For the results given by equations~\eqref{valorH0suma}
and~\eqref{valorH0prod}, 
$\Omega_{eff}$ has the following values:

\begin{equation}
    \Omega_\text{eff}=5.630465^{+1.903135}_{-1.530342}\quad \mbox{and} \quad
    \Omega_\text{eff}=5.699477^{+0.482638}_{-0.449577},
    \label{efectiva}
\end{equation}

\noindent for the weak and the strong curvature-matter coupling models respectively. These expressions
are in great agreement with the empiric relation: $H_0=2\pi a_0/c$, reported by \citet{Milgrom1}.  Note that the ``correct'' definition 
of the density parameter at the present epoch for each model would have to be: \( \Omega_\text{correct} =  a_0 \Omega_\text{eff} / 
c H_0 \), which has a value equals to \( 1 \) according to equation~\eqref{effective}.

  The fact that our obtained values for $H_0$ are within the range of the current inferred values, give us the
certainty that the calibrated values of our three free parameters $q_0$, $j_0$ and $\Omega_0$ are reliable.  

Since both weak and strong curvature-matter coupling  models presented in this article have the same number of parameters, then the best fitting model is the one for
which the SSR is minimal.  
From the results of Table~\ref{table02} and~\ref{table03} it follows that the best fitting model is the one with weak curvature-matter coupling.  However the advantage between 
one and the other is minimal and in principle both reproduce quite well SN~Ia observations.

  In summary, if one requires to build a local or non-local relativistic theory of MOND in 
the pure metric formalism it is necessary to introduce curvature-matter couplings into the 
action.  The infinite possibilities that arise with this choice need to be 
compared with more astronomical and cosmological observations in order to find a ``unique'' correct theory.
Also, an equivalent Parametrised
Post-Newtonian formalism needs to be developed at least in the deep-MOND regime and 
more comparisons with cosmological phenomenology and are to be performed.  
The constraints for possible positive candidates can be obtained using data
from by the fundamental plane of elliptical
galaxies.  In these pressure supported systems, the pure \( f(R) \) techniques
developed by~\citet{jovanovic16,capozziello17} represent a useful tool to 
be generalised for the curvature-matter couplings reported in this article.  
In this respect, using Yukawa-like solutions such as the ones reported 
by~\citet{capozziello20} may provide a good way to deal with the required
constrains.

\section*{Acknowledgements}
This work was supported
by DGAPA-UNAM (IN112019) and CONACyT 
(CB-2014-01 No.~240512 and Ciencia de Frontera No.~304001) grants. 
EB, TB and SM acknowledge economic support from CONACyT 
(517586, 64634 and 26344).

\bibliographystyle{apsrev}
\bibliography{barrientos_mendoza}

\appendix 
\label{ape}
\section{MOND as the weak-field limit of $F(\chi,\xi)=\chi^{-3}+\xi^{-2}$}

In Section~\ref{cosmo}, we proposed the function $F(\chi,\xi)=\chi^\gamma
+\xi^\beta$ and in Section~\ref{ajuste} we fit data for $\gamma=-3$ and
$\beta=-2$. These values are justified because in the weak-field limit
of the theory a MONDian behaviour for the acceleration is obtained.

Let's start with the field equations of the theory given
by equation~\eqref{field.eqs.}. Taking the trace, substituting
$F(\chi,\xi)=\chi^\gamma +\xi^\beta$ and using the definitions of $\xi$
and $\chi$, we obtain the following equation:

\begin{equation}
    (\gamma-2)\alpha^\gamma R^\gamma+3\alpha^\gamma \gamma\Delta R^{\gamma-1}=\frac{2(1-\beta)}{\lambda^\beta}\lmatt^\beta+\frac{\beta}{2\lambda^\beta}T\lmatt^{\beta-1},
    \label{traza.mond}
\end{equation}

\noindent for dust $\lmatt=T$, then the previous equation turns into:

\begin{equation}
    \overbrace{(\gamma-2)\alpha^\gamma R^\gamma}^{{\cal{O}}(2\gamma)}+\overbrace{3\alpha^\gamma \gamma\Delta R^{\gamma-1}}^{{\cal{O}}(2(\gamma+1))}=\overbrace{\left(2-\frac{3\beta}{2}\right)\left(\frac{\lmatt}{\lambda}\right)^\beta}^{{\cal{O}}(2\beta)},
    \label{traza.mond.dust}
\end{equation}

\noindent where the order in $c^{-1}$ is shown. Since $2(\gamma+1)$
is always greater than $2\gamma$, the following choice seems natural:

\begin{equation}
    \gamma+1 =\beta.
    \label{valores}
\end{equation}

Therefore, the equation for the dominant order in $c$ is:

\begin{equation}
  3\alpha^\gamma \gamma\Delta R^{\gamma-1}=\left(\frac{1-3\gamma}{2}\right)\left(\frac{\lmatt}{\lambda}\right)^{\gamma+1}.
    \label{dominante}
\end{equation}

The weak-field limit for the previous equations at order of magnitude is given by: 

\begin{equation}
    a \sim \frac{(GM)^{(\gamma+1)/(\gamma-1)}}{a_0^{2/(\gamma-1)}}r^{-2(\gamma+1)/(\gamma-1)},
    \label{Mond}
\end{equation}

\noindent in order to recover the MONDian acceleration $a=(GMa_0)^{1/2} r^{-1}$
the following value for $\gamma$ is found:

\begin{equation}
    \gamma=-3 \qquad \rightarrow \qquad \beta=-2.
    \label{valores2}
\end{equation}

 Now, we will perform an analogous approach as the one made by \citet{barrientos18}
in order to find the values of the constants $k$ ans $k'$. Direct substitution
of the parameters $\gamma$ and $\beta$, as well as of $R=-2\nabla^2\phi/c^2$ in 
equation \eqref{traza.mond.dust} for the order of interest yields to:

\begin{equation}
    -\frac{9(a_0G)^2}{5\cdot 2^4 k^3 k'^2}\nabla^2(\nabla^2\phi)^{-4}=\frac{1}{\rho},
    \label{intermedio}
\end{equation}

\noindent where the matter Lagrangian for dust has been employed. For a point-mass source
in spherical symmetry:

\begin{equation}
    \rho=\frac{M \delta(r)}{4\pi r^2}, \quad \mbox{and} \quad \nabla^2\phi=\frac{1}{r^2}\frac{\text{d}}{\text{d}r}\left(r^2\frac{\text{d}\phi}{\text{d}r}\right).
\label{esfericas}
\end{equation}

With these assumptions, equation \eqref{intermedio} turns into:

\begin{equation}
    -\frac{9}{5\cdot 2^4 k^3 k'^2}\left(\frac{Ma_0G}{4\pi}\right)^2\frac{\text{d}}{\text{d}r}\left[r^2
    \frac{\text{d}}{\text{d}r}\left(\nabla^2\phi\right)^{-4}\right]=\frac{r^6}{\left[\delta(r)\right]^3}\delta(r),
    \label{intermedio2}
\end{equation}

\noindent integrating with respect to $r$ and using the Dirac's delta function as: $\delta(r=0)=\lim_{r\rightarrow 0}(2\pi r)^{-1}$
\citep{delta}, the previous equation is given by:

\begin{equation}
    -\frac{9}{5\cdot 2^{11} k^3 k'^2 \pi^5}(Ma_0G)^2\frac{\text{d}}{\text{d}r}(\nabla^2\phi)^{-4}=r^7.
    \label{intermedio3}
\end{equation}

Performing two more integrations with respect to $r$ and since the acceleration is defined as: 
$a=-\nabla\phi$, the final expression for the acceleration is: 

\begin{equation}
    -\left[-\frac{72}{5\cdot 2^{11} k^3 k'^2 \pi^5}\right]^{1/4}\frac{(Ma_0G)^{1/2}}{r}=a.
    \label{intermedio4}
\end{equation}

In order to obtain the MONDian expression for acceleration: $a=-\sqrt{Ga_0M}/r$, the following 
relation for $k$ and $k'$ holds:

\begin{equation}
    k^3k'^2=-\frac{72}{5\cdot 2^{11}\pi^5}\approx-2.3\times 10^{-5}.
    \label{ks}
\end{equation}

It is quite important to stress that the calculations made in this
Appendix are only relevant for computing the coupling constants in
equation~\eqref{ks}.  As stated in the last paragraph of Section~\ref{generalities}, the weak curvature-matter coupling 
used in this article is only valid for no-vacuum cases, i.e. when
\( \rho \neq 0 \).  In fact, equations~\eqref{intermedio3} and~\eqref{intermedio4}  are  only valid
at the origin of coordinates, i.e. when \( \rho \rightarrow \infty \) 
and should 
never be extended further beyond that limit for the case of a point mass 
source.

\end{document}